%Paper: hep-th/9308158
%From: Ivan Kostov <kostov@amoco.saclay.cea.fr>
%Date: Thu, 2 Sep 1993 17:07:30 +0000

\input harvmac
%\draftmode
\def\u{{\bf u}}
\def\U{{\bf U}}
\def\V{{\bf V}}
\def\H{{\bf H}}
\def\v{{\bf v}}
\def\h{{\bf h}}
\def\CS{{\cal S}}

\def\CN{{\cal N}}

\font\cmss=cmss10 \font\cmsss=cmss10 at 7pt
\def\IZ{\relax\ifmmode\lrefhchoice
{\hbox{\cmss Z\kern-.4em Z}}{\hbox{\cmss Z\kern-.4em Z}}
{\lower.9pt\hbox{\cmsss Z\kern-.4em Z}}
{\lower1.2pt\hbox{\cmsss Z\kern-.4em Z}}\else{\cmss Z\kern-.4em Z}}
\def\np{Nucl. Phys. }
\def\pl{Phys. Lett. }
\def\pr{Phys. Rev. }

\def\CF{{\cal F}}

\def \CM{{\cal M}}

\def \l{\ell}

\def \sinh{{\rm sinh}}

\def\I{{\bf I}}
\def\CD {{\cal D}}

\def\CL {{\cal L}}

\def\p {\partial}
\def\CS {{\cal S}}
\font\cmss=cmss10 \font\cmsss=cmss10 at 7pt
\def\IZ{\relax\ifmmode\mathchoice
{\hbox{\cmss Z\kern-.4em Z}}{\hbox{\cmss Z\kern-.4em Z}}
{\lower.9pt\hbox{\cmsss Z\kern-.4em Z}}
{\lower1.2pt\hbox{\cmsss Z\kern-.4em Z}}\else{\cmss Z\kern-.4em Z}}

\def\l {\ell }

\def\a{\alpha}
\def\ta{\tilde \alpha}

\def\L {\Lambda }

\def\tr{ {\rm tr}}
\def\p {\partial}
\def\CS {{\cal S}}
\def\CL {{\cal L}}
\def\CF {{\cal F}}
\def\CZ{{\cal Z}}

\def\CD {{\cal D}}
\def\pl{{\it Phys. Lett. } }
\def\np{{\it Nucl. Phys.} }
\def\pr{{\it Phys. Rev.} }

\def\R{\relax{\rm I\kern-.18em R}}
\font\cmss=cmss10 \font\cmsss=cmss10 at 7pt
\def\Z{\relax\ifmmode\mathchoice
{\hbox{\cmss Z\kern-.4em Z}}{\hbox{\cmss Z\kern-.4em Z}}
{\lower.9pt\hbox{\cmsss Z\kern-.4em Z}}
{\lower1.2pt\hbox{\cmsss Z\kern-.4em Z}}\else{\cmss Z\kern-.4em Z}}

\lref\notrn{A. M. Polyakov, \pl 72B (1978) 447;
J.-M. Drouffe, \pr D18 (1978) 1174;
L. Susskind, \pr D20 (1979) 2610;
M. Nauenberg and D. Toussaint, \np B190 (1981) 288;
P. Menotti and E. Onofri, \np B190 (1984) 288;
C.B. Lang, P. Salomonson and B.S. Skagerstam, \pl 107B (1981) 211,
\np B190 (1981) 337}.
 \lref\ros{P. Rossi, {\it Ann. Phys.}
132 (1981) 463}
\lref\mm{Yu. M. Makeenko and A. A. Migdal, \np B188 (1981) 269,
A. A. Migdal, \pr 102 (1983) 201}
\lref\tho{G. 't Hooft, \np B72 (1974) 461}.

\lref\rus{B. Rusakov, {\it Mod. Phys. Lett.} A5 (1990) 693}
\lref\gt{ D. Gross, \np B400 (1993) 161;
 D. Gross and W. Taylor IV, \np B400 (1993) 181}
 \lref\dg{ M. Douglas, Preprint RU-93-13 (NSF-ITP-93-39);
J. Minahan and A. Polychronakos, hep-th/9303153}
\lref\dgk{ M. Douglas and V. Kazakov, preprint LPTENS-93/20}

\Title{T93/079}
%{\vbox{\baselineskip12pt\hbox
%{SPhT/93-050}\hbox{ }}
{\vbox{\centerline
{$U(N)$ Gauge Theory and Lattice Strings   }
\vskip2pt
\centerline{
 }}}

\vskip4pt

\centerline{Ivan K. Kostov \footnote{$ ^\ast $}{on leave of absence
from the Institute for Nuclear Research and Nuclear Energy,
Boulevard Tsarigradsko Chauss\'ee 72, BG-1784 Sofia, Bulgaria}
\footnote{$^{\# }$}{(kostov@amoco.saclay.cea.fr)}
}
\centerline{{\it Service de Physique Th\'eorique
\footnote{$ ^\dagger$}{Laboratoire de la Direction des Sciences
de la Mati\`ere du Comissariat \`a l'Energie Atomique} de Saclay
%}
%\centerline{
CE-Saclay, F-91191 Gif-Sur-Yvette, France}}

\vskip .3in

\baselineskip12pt{
We explain, in a slightly modified form, an old construction
allowing to reformulate the  $U(N)$ gauge theory defined on a
$D$-dimensional lattice
as a theory of lattice strings (a statistical model of  random surfaces).
The world surface of the lattice string is  allowed to have pointlike
 singularities (branch points)
located not only at the sites of the lattice, but also on
 its  links and plaquettes.
The strings become  noninteracting
when  $N\to\infty$.  In this limit the statistical weight a world
surface is given by exp[ $-$ area] times a product of local factors
associated with the branch points.
In $D=4$ dimensions the gauge theory has a nondeconfining first order
phase transition dividing the weak and strong coupling phase.
{}From the point of view of the string theory the weak coupling phase
is expected to be characterized by spontaneous creation of ``windows'' on the
world sheet of the string.

}

\vskip 1cm
\leftline{Talk delivered at the {\it Workshop on string theory, gauge
 theory
and quantum gravity},}

\leftline{ 28-29 April 1993, Trieste, Italy }
\smallskip
%\draft
\Date{ August 1993 }

\baselineskip=14pt plus 2pt minus 2pt

\newsec{Introduction}
The concept of gluons as elementary excitations in QCD is
adequate  only at distances much less than the correlation
length of the theory (the typical hadron radius).
  Because of the dimensional transmutation, the
effective  gauge
coupling constant seizes to be a small parameter in the
infrared domain and the standard perturbation theory collapses
 at distances comparable
with the hadron radius.

An idea which is perhaps  as old as QCD is
 that  infrared-stable
elementary excitations are described by fluctuating surfaces or strings.
The Wilson's strong coupling expansion
\ref\wils{K. Wilson, \pr D 10 (1974) 2445; A. M. Polyakov, $unpublished$}
and the $1/N$ expansion of 't Hooft
\tho\
 are the two basic arguments in favour
of the possibility of such a description.

{}From the point of view of the $1/N$ expansion,
the $U(N)$
gauge theory  resembles  a theory of strings
with interaction constant $1/N$ and its  multicolor limit
 $N\to \infty$ is expected to
be described by some kind of a noninteracting string.
The formation of strings at large distances can be  seen as
 a condensation of
dense planar diagrams in the continuum theory (fig. 1).
However, the  only unambiguous evidence for the relevance of
strings in the perturbation theory  is the topology of the Feynman diagrams.
Because of the divergencies and the need of gauge fixing, it is not easy to
make this statement more rigorous.

\vskip 4 cm

\centerline{Figure 1. A dence planar diagram for the Wilson
loop average
 in the multicolour QCD}
\bigskip

On the other hand, there is the
  strong coupling expansion
in  lattice gauge theories \wils.  The gauge theory
 is discretized so that
   the gauge field is associated with the
links $\l = <xx'>$  and the action is a sum over the elementary
squares (plaquettes)   of the lattice.
{}From the point of view of the strong coupling expansion
the lattice gauge theory is  statistical mechanics of
random surfaces with contact (self)interactions.
In the lowest orders of the inverse coupling $\beta = 1/\lambda$,
the Wilson loop average is given by the sum over lattice surfaces spanning
the loop.
In high orders, the dominant contribution comes from overlapping surfaces
subjected to strong contact interactions and forming new topologies,
and the original string picture is lost.

Some years ago, it has been suggested that the large $N$ limit of
the strong coupling expansion might correspond to a  theory
of noninteracting surfaces.
In order to study this possibility, the most commonly used
 character expansion
was not the most efficient  mathematical  formalism
 because it produces diagrams
for the partition function and not directly for the free energy.
In spite of the considerable efforts
\ref\workk{I. Bars, {\it Journ. Math. Phys.} 21 (1980) 2678;
S. Samuel,   {\it Journ. Math. Phys.} 21 (1980) 2695;
R.C. Brower and M. Nauenberg, \np B180 (1981) 221;
D. Weingarten, \pl 90B (1980) 280;
D. Foerster, \np B170 (1980) 107}
no evidence for a free string picture was obtained
\foot{Nevertheless, very interesting results concerning
 ${\rm QCD}_2$ have been obtained
in a series of recent papers \refs{\rus {--} \dgk}
using the method of characters.}.

An alternative formalism was then suggested by V. Kazakov
\ref\kaza{V.Kazakov, \pl 128B (1983) 316, JETP (Russian
edition) 85 (1983) 1887}.
The idea of  \kaza , borrowed from the Stanley's
 solution \ref\st{H.E. Stanley,
\pr 176 (1968) 718} of the
large-$N$ vector model,
  was to reformulate the $U(N)$ lattice gauge theory
as a kind of  Weingarten theory with additional interactions.
The Weingarten model \ref\wein{D. Weingarten,
\pl 90B (1980)285}
is a simple (but
rather pathological)
gauge theory which is equivalent to a  model of noninteracting
 random surfaces in the large $N$ limit.
%It can be described as a $U(N)$ pure lattice gauge theory,
% with the unitary measure replaced by a Gaussian measure in the
%space of complex $N\times N$ matrix field variables.
In the $U(N)$ lattice gauge theory one can
 impose the unitarity   of the link variable
$\U_{\l}$ by means of a Lagrange multiplier and  replace the Haar measure on
the
$U(N)$ group by a Gaussian  measure over complex matrices.
 The idea of \kaza\ has been accomplished in refs.
\ref\obrz{K.H. O'Brien and J.-B. Zuber, \np B253 (1985) 621,
\pl 144B (1984) 407}, \ref\kos{I. Kostov, \pl
138B (1984) 191, 147B (1984) 445}.

In ref. \ref\ivn{I. Kostov, \np B265 (1986) 223}
we suggested the possibility that the strong coupling lattice surfaces
and planar Feynman-'t Hooft  diagrams appear as two
 extreme cases of
the same perturbative expansion.
This means that the strong coupling lattice string is perhaps not completely
irrelevant to the continuum limit of QCD.

Here we explain the construction of ref. \ivn\ using the standard formalism
of the mean field analysis in gauge theories
 \ref\jb{J.-M. Drouffe and J.-B. Zuber, {\it Phys. Rep.}
102, Nos. 1,2 (1983)1-119, section 4}.
Our  random surface $Ansatz$ will be defined for any choice of the
one-plaquette action. In $D=2$ dimensions it reproduces the continuum
limit of the gauge theory in spite of eventual
Gross-Witten transition
\ref\gw{D. Gross and E. Witten, \pr D21 (1980) 446}
 separating the strong and weak coupling
domains.
However, the first-order phase transition in $D=4$ dimensions
is likely to persist and the sum over surfaces has to
be modified in the weak coupling phase by allowing windows (free boundaries)
to appear on the world sheet.
In the continuum limit the world sheet of the string is tore into a
network of thin strips which can be interpreted as gluon propagators.

We start the story with a review of
the Stanley's solution \st\
of the
 large-$N$ vector model which may help the reader to
understand the logic of our construction before dipping
into the   technical details.
The analogy with the vector model can help us  to
 get some more intuition  and avoid  the rudest
conceptual errors in the case of the gauge theory.

\newsec{The $U(N)$ vector model}

The fluctuating field in this model
  is an   $N$-component complex  vector $\u_x=(u_x^1,...,u_x^N)$
defined at the  points $x$ of the $D$-dimensional hypercubical
lattice
and having unit norm
\eqn\nnor{\u^*_x \cdot
\u_x \equiv \sum _{i=1}^{N} |u^i_x|^2 =1}
We assume  a nearest neighbour interaction  $\beta \u^*_x \cdot
\u_y $ associated with
the links $\l = <xy>$ of the lattice.
Then the measure
\eqn\unim{[d\u]= \prod_{i=1}^N d\bar u^i du^i
\ \delta ( \sum _{i=1}^{N} |u^i_x|^2  -1)
= d\u^*d\u \ \delta (\u^*\cdot \u -1) }
 and the interaction are invariant under global $U(N)$
transformations
\eqn\trrt{\u_x \to \U  \u_x; \ \ \U\in U(N)}
The partition function
\eqn\prtf{\CZ =e^{N\CF} = \int \prod _{x}[d\u_x]
\prod _{\l = <xy>} e^{N\beta \u^*_x \cdot \u_{y}}}
can be calculated order by order in the inverse coupling constant $\beta$
by  expanding the exponential as a series of monomials
(strong coupling expansion).
This leads to a diagram technique which in the limit $N\to \infty$
involves tree-like clusters of loops and can be  summed up
explicitly in all orders.

Alternatively one can  transform, by introducing auxiliary fields, the
integral over the original vector field into a duable Gaussian integral
and find an effective
field theory  in which the number $1/N$ plays the r\^ole
of a Planck constant.

 Introducing a new complex vector field
$\h=(h^i,\  i=1,2,...,N)$ we represent the unimodular measure as a flat measure
for the unconstrained complex vector field $\v=(v^i, i=1,...,N)$.
We insert the exponential representation of the $\delta$-function
\eqn\ddd{\delta (u^j -v^j)= {1 \over 2\pi i}
\int_{-i\infty}^{i\infty}dh_j e^{\bar h_j(u^j-v^j)}}
in the definition \prtf\ of the partition function for every component
$u^j_x$ of the vector field.
Here $\h_{\l}$ and $\v_{\l}$ are unrestricted complex vectors
on which the integration is performed with a flat measure
\eqn\mmemmm{d\v = \prod_{j=1}^N dv^j d\bar v^j,\ \ \
d\h = \prod_{j=1}^N dh^j d\bar h^j}
Denoting by
\eqn\ffff{e^{NF(\h^* \cdot \h)}= \int [d\u]e^{N[\u^* \cdot
 \h+\h^*\cdot \u]}}
the one-site mean-field integral we write the partition function \prtf\
in the form
\eqn\pppar{
\CZ=e^{N\CF}= \int \prod _{x }\Big(
d\h_{x}d\v_{x}
e^{-N (\h^*_x \cdot \v_x + \h_x \cdot \v_x^*
)}\Big)
\prod_{x} e^{N
F(\h_x)} \prod _{\l = <xx'>}e^{N\beta \v_x\cdot\v_x'}}

The potential $F(\a)$ is determined by the Ward identity
$(\sum_i N^{-2} \p/\p h^i\p/\p h^*_i - 1)e^{NF}=0$ or, in
 terms of the $\a$-field,
\eqn\wwwar{({1 \over N}{\p \over \p \a}+\a{1 \over N^2} {\p^2 \over \p \a^2}-1)
e^{NF(\a)}=0}
which is solved, in the large $N$ limit, by
\eqn\aaaa{\eqalign{
{\p F\over \p \a}&={\sqrt{1+4\a}-1 \over 2\a}\cr
F(\a) &=\sqrt{1+4\a}-1-\ln {1+\sqrt{1+4\a} \over 2}=\sum_{n=1}^{\infty}
f_{n}{\alpha^n \over n}
\cr}}

Using the integral representation
\eqn\iiii{e^{NF(\h^*\cdot \h)}=\int d\a d\ta e^{N[\ta (\h^*\cdot \h-\a)+
F(\a)]}}
we finally write the partition function
 in a form allowing to perform exactly the integral
over the vector fields
\eqn\npff{\eqalign{
\CZ=e^{N\CF}=&\int \prod_{x}d\a_x d\ta_x\ e^{N[F(\a_x)-\ta_x\a_x]}
 d\v_x d\h_x
e^{N[-\v^*_x\cdot \h_x-\h^*_x\cdot \v_x
]}\cr
&\prod_x  e^{N[\ta _x \h ^*_x\cdot
\h_x ]}
\prod_{\l=<xx'>}e^{N\beta \v^*_x\cdot \v_{x'}}\cr}
}
Denoting by $\hat C$
the connectivity matrix of the $D$-dimensional hypercubical   lattice
\eqn\cccac{\hat C=\sum_x \sum_{\mu =1}^{D}
(e^{\p/\p x_\mu}+e^{-\p/\p x_\mu})}
we can write the  result of the   gaussian integration over  the $\v,\h$ fields
as
\eqn\zzz{e^{N\CF}=\int \prod_{x}d\ta_xd\a_x e^{N[
 F(\a_x ) - \ta_x\a_x]}\ \  e^{N\CF_{0}[\ta]} \equiv
\langle e^{N\CF_0[\ta]}\rangle_{\ta}}
where
\eqn\ffrf{\CF_{0}[\ta]=-\sum_{x} \tr \  \log (1- \beta \ta \hat C)}
is the sum over vacuum loops of the vector fields.

Formally the sum over vacuum loops can be obtained by
expanding the exponential in the last line of \npff\ as a sum of
monomials and applying all possible Wick contractions
\eqn\wwik{\langle \h^{*i}_x \v_{yj}\rangle = \delta_{x,y} \delta_{j}^{i},
\langle \v^{*i}_x \h_{yj}\rangle = \delta_{x,y} \delta_{j}^{i}}
The vertex $\ta_x \h^*_x\cdot \h_x$  is represented by a node at the point $x$
and contributes a factor $\ta_x$ to the weight of the loop. The
vertex $\beta \v^*_x\v_y$ is represented by a segment of the loop
 covering the link
$\l =<xy>$ and contributes a factor $\beta$.
 The weight of each loop is therefore a
 product of local factors  $\beta$ and $\ta_x$
associated with its nodes and links, correspondingly.
Denoting by $\Gamma$ any closed loop on the lattice,
 the r.h.s. of \ffrf\ can be written as
\eqn\spts{\CF_0[h^*]=\sum_{\Gamma}\beta ^{\#{{\rm links}}}
\prod _{x\in \Gamma}\ta_x }

The action for the fields  $\ta , \a $ is proportional to $N$ and in the
limit $N\to \infty$ these fields freeze at their
 expectation values determined by the classical equations of motion
\eqn\eqnm{\ta ={\p F(\a)\over \p \a}={\sqrt{1+4\a}-1 \over 2\a}}
\eqn\eqnmm{\a={\p \CF_0(\ta)\over \p \ta}={\beta \hat C
\over 1- \ta \beta \hat C}}
or, equivalently,
\eqn\ssas{\ta +\a \ta ^2 =1, \ \ {\ta  \over 1-\beta \ta \hat C}=1}
The last equation has a simple geometrical interpretation in terms of
a random walk on the lattice.
 It can be obtained also directly from the
unimodularity condition $\u^*_x\cdot \u_x=1$.
Consider the two-point Greens function $G_{xy}=\langle \u^*_x\cdot
\u_y\rangle$.
Repeating the same steps as for the partition function we find
\eqn\mnmbm{G_{xy}=\langle \v^*_x\cdot \v_y\rangle = \langle x|{\ta \over
1-\beta \hat C}|y \rangle=\sum _{\Gamma_{xy}}\ta ^{\# {\rm nodes}}
\beta^{\#{\rm links}}}
Therefore the second of the equations \ssas\ is just the condition
\eqn\uuuu{G_{xx}=1.}
We see that the original vector field $\u_x$ acquires a longitudinal
degree of freedom and instead of $2N-1$  we have $2N$ degrees of freedom
at each point $x$.
These describe $N$ identical noninteracting particles with
 dynamically generated mass.
The constraint $\u^*_x\cdot
\u_x=1$ is imposed on dynamical level by the condition
\uuuu\  which determines the mass gap of the model.
 After diagonalysing
 the
connectivity matrix we write it as an integral in the momentum space
\eqn\tttr{\int_{-\pi}^{\pi} {1 \over m^2+\sum_{\mu =1}^D[2\sin (p_{\mu}/2)]^2}
=\beta}
\eqn\mmmas{m^2= {1 \over \ta \beta} -2D}
The continuum limit (infinite correlation length) is achieved when
 $m^2\to 0$, i.e.,  $ 2D\beta \ta =1$.

Let us examine more closely this equation depending on the
dimension $D$.

\subsec{$D\le 2$}

When
 $D\le 2$, the
integral is infrared divergent when $m\to 0$ and gives
\eqn\divv{ \beta =\int {d^Dp \over (2\pi)^D} {1 \over p^2+m^2}
= m^{D-2}, \ \ D<2}

In the marginal case $D=2$ we find the asymptotic freedom law for the
dependence
of the
renormalized coupling $\beta = \beta (m)$ as a function of the
correlation length $1/m$
\eqn\assf{\beta = {1 \over4\pi} \ln {1 \over m^2}}

\subsec{$D>2$}

Above two dimensions the integral is convergent at $m\to 0$ and
the limit of zero mass is achieved at finite value of $\beta = \beta _c$.

  In the weak coupling phase $\beta >\beta _c$ the equations of
motion  \tttr\ have no solution which means that something in our
construction is wrong.
This contradiction is resolved if we admit that the
vector fields acquire nonzero vacuum expectation values
 $\langle \v_x \rangle =\v_{{\rm cl}}$,  $\langle \h_x \rangle =\h_{{\rm cl}}$
determined by the saddle-point equations
\eqn\axi{ \h_{{\rm cl}}=\beta \hat C \v_{{\rm cl}}= 2D\beta
\v_{{\rm cl}},\ \  \v_{{\rm cl}}= \ta  \h_{{\rm cl}}}
which are solved by  $     \h_{{\rm cl}}= \v_{{\rm cl}}=0$
( strong coupling  $U(N)$-symmetric phase) or
 by  $m^2= 1- 2D\beta \ta =0$ with
    $ \h_{{\rm cl}}, \v_{{\rm cl}} \ne 0$ ( weak coupling
 phase with spontaneously
broken $U(N)$ symmetry).
 Geometrically the weak coupling phase is characterized by
possibility of the
  the random walk to
 break into two pieces.
Then the two-point Greens function is a sum of  connected and  disconnected
parts
\eqn\gggr{G_{xy}=
\v_{{\rm cl}}^2+\langle x|{\ta \over 1- \ta \beta \hat C}|y\rangle}
The unimodularity condition now reads
\eqn\rrr{
\v_{{\rm cl}} ^2
 +{1 \over \beta}\int {d^Dp \over (2\pi)^D} {1\over
\sum_{\mu=1}^D(2\sin (p_{\mu}/2))^2}
=\v_{{\rm cl}} ^2
+{\beta _c  \over \beta}=1}
Thus in the weak coupling phase there is a spontaneous magnetization
$\v_{{\rm cl}}
=\sqrt{1-\beta_c/\beta}<1$ and  $2N-1$ Goldstone excitations
with zero mass due to the spontaneous breaking of the $U(N)$ symmetry.

The critical value $\beta _c$ is related to the probability $\Pi_0$
that a Brownian walk returns to the origin
\ref\id{J.M. Drouffe and C. Itzykson, {\it Statistical Field Theory},
Cambridge University Press 1989}
 \eqn\ppi{2D\beta_c ={1 \over 1- \Pi_0}}
and its large behaviour is given by
\eqn\ldd{2D\beta_c=1+{1 \over 2D}+3\Bigg( {1 \over 2d}\Bigg)^2+...}
The series diverges at $D=2$ and as $D\to 2 , \beta_c \to \infty$
\eqn\eiei{2D\beta_c \sim {2 \over \pi (D-2)}}
and below the lower critical dimension $D=2$ no transition occurs.
In $D<2$ there exists only a high temperature symmetric phase.

\subsec{$D=1$}

The $D=1$ vector model
has no dynamical degrees of freedom and is  analogous  to the
 $D=2$ gauge theory.

In terms of  the weak coupling parameter  $\lambda$ related to $\beta$ by
$2/\beta = \sinh (\lambda /2)$ we find
\eqn\pppao{\ta = \tanh {\lambda \over 2}  ,
\  \  \beta = {1 \over 2 \sinh \lambda /2}}
 and the two-point correlator is given by
\eqn\dimii{G_{xy}=e^{-(\lambda/2) |x-y|}}
The two-point correlator can be interpreted either as the sum over all random
walks (with backtrackings) connecting the points $x$ and $y$,
 or as the only minimal path (without backtrackings)
  connecting the two points,
with a weight factor $G_1=e^{-\lambda /2}$ associated with each of its links.
In the first interpretation the mean value of $\ta$
is tuned so that
it compensates
completely the entropy due to backtrackings.
In the case of the gauge theory
the analog of the backtrackings
are the folds of  the world sheet of the string.
We shall see that the $D=2$ gauge theory can be formulated in terms of
minimal nonfolding surfaces.

\newsec{The $U(N)$ gauge theory}

In the Wilson-Polyakov formulation of the $U(N)$
lattice theory \wils\
the independent fluctuating variable is a unitary matrix
${\bf U_{\l}}=\{(U_{\l})_i^j, i,j=1,...,N\}$
\eqn\unitt{{\bf U}_{\l}^{\dag}{\bf U_{\l}}={\bf I}} associated with the
oriented links $\l = <xy>$ of the hypercubical lattice\foot{We will use the
notation $\l^{-1}=<yx>$ for the link obtained from $\l = <xy>$
by reversing the orientation.
More generally, if  $\Gamma_1$ and $  \Gamma_2$
  are two oriented lines, the product  $\Gamma_1 \Gamma_2$ is defined,
in case it exists, as the
line obtained by identifying the end of   $\Gamma_1$ with the beginning
of $ \Gamma_2$.}.
The measure and the interaction are invariant with respect to local
gauge transformations
\eqn\gggee{{\bf U}_{<xy>} \to {\bf U}_x  \U_{<xy>}\U_{y}^{\dag}}
The amplitude  $\U(\Gamma)$ associated with the parallel
 transport along the loop
$\Gamma=\{\l_1\l_2 ... \l_n\}$ is given by the product of the link variables
\eqn\spomn{\U(\Gamma)= \U_{\l_1} \U_{\l_2}...\U_{\l_n}}
The action in the gauge theory is a functional on the loop fields \spomn .
The simplest nontrivial
 loop  $\Gamma$ is the boundary $\p p$ of a
 $plaquette$ $p$, an elementary square
on the hypercubical lattice. In what follows we will denote by $\U_p$
the corresponding loop variable,
the ordered product of the  4 link variables along the boundary of the
plaquette $p$.
Let us denote by
$S_{\lambda}(\U) $ the  one-plaquette action
where $\lambda$ is the coupling constant of the lattice gauge theory.
Then the  partition function is defined by the integral over all link variables
\eqn\discr{
Z =e^{N^2 \CF}= \int _{\l \in \CS} [d\U_{\l}]
\prod_{p }e^{S_{\lambda}(\U_ p)}}
where
\eqn\measr{[d\U]= d\U d\U^{\dag} \delta(\U^{\dag}\U-\I)
\equiv  \prod_{i,j=1}^{N} dU^j_id (U^{\dag})^j_i \delta \Big(
\sum _k  (U^{\dag})_i^k U_k^j -\delta _i^j\Big)
}
 is the invariant measure on the group $U(N)$.
By
 $\CF$ we denoted the  free energy per degree of freedom.

The one-plaquette action is subjected to
the following two requirements.
First, it should be a  real function
defined on the
conjugacy classes of the group $U(N)$.
This means that it can be expanded in the characters of the irreducible
representations of the group $U(N)$
\eqn\chpo{e^{N^2S_{\lambda}(\U)}= \sum_{-\infty
\le n_1\le ...\le n_N\le \infty}
\chi_{\vec n}(\I)\chi_{\vec n}(\U) e^{N^2S_{\lambda}^{\vec n} }}
The character of the representation with signature $\vec n
=\{ n_1,...,n_N\}$ depends on
the eigenvalues $u_1,...,u_N$ of the unitary matrix $\U$ as
\eqn\chhr{\chi_{\vec n} (\U)= {\det_{ik} (u_i^{n_k+k-1}) \over \det _{ik}
(u^{k-1}_i) }.}
Second, we assume that the continuum limit is achieved when $\lambda \to 0$
and in this limit
\eqn\llla{S_{\lambda}(\U) \to -{1 \over 2 } {\tr \over N} {\bf A}^2
, \ \ \  \U=e^{i\sqrt{\lambda} {\bf A}}.}
The Fourier image of the action in this limit is proportional to  the second
Cazimir operator
\eqn\rsrsr{S_{\lambda}^{\vec n} = {\lambda \over 2N} C_2(R) =
{\lambda \over 2N} \Big( \sum _{i=1}^N n_i^2
+\sum _{i<j} (n_i-n_j)\Big)}
If we assume that eq. \rsrsr\ remains true also for finite values of $\lambda$,
the corresponding one-plaquette action is known as {\it heat kernel action}
\notrn ,  $S^{HK}_{\lambda}(\U)$.
The exponential of this action
 is a solution of the  heat kernel equation on the group manifold
\eqn\hhhhk{ \big[ 2{\p \over \p\lambda}
+ (\U\p / \p \U)^2\big]e^{N^2S_{\lambda}^{HK}(\U)}=0}
The heat kernel action has the nice property of reproducing itself
after  group integration \ros
\eqn\reppo{ \int [d\U] e^{N^2S^{HK}_{\lambda_1}(\U_1\U)}
 e^{N^2S^{HK}_{\lambda_2}(\U_2\U^{-1})}
= e^{N^2S^{HK}_{\lambda_1 +\lambda_2}(\U_1 \U_2)}}

Another simple choice is the {\it  Wilson action} \wils\
 $ S_{\lambda}^{\rm Wils} (\U) = {1 \over 2 \lambda} {\tr
\over N}( \U +\U^{\dag})$.
Its Fourier image is given by two different analytic expressions
in the limit $N\to \infty$, depending on the value of the coupling
constant.
For example, for the fundamental representation $\vec n_f =
[0,0,...,0,1]$
\eqn\kkkdk{S_{\lambda}^{\vec n_f}=
\cases{
{1 \over 2\lambda } , &if $\lambda >1$; \cr
1-{1 \over 2}\lambda , & if   $\lambda <1$ \cr}}
The nonanalyticity at the point $\lambda =1$  known as Gross-Witten
phase transition  \gw\
is due to the fact that the integral for the
 inverse Fourier transform
\eqn\hahaha{e^{N^2S_{\lambda}^{\vec n}}={1
 \over \chi_{\vec n}(\I)} \int [d\U] e^{N^2S_{\lambda}(\U)} \chi _{\vec n}(\U)}
 is saturated by
the vicinity of a saddle point and the saddle-point solution can have two
analytic branches.
In the strong coupling domain $\lambda >1$  the eigenvalues
of the unitary matrix $\U$ are distributed all along the unit circle,
and  in the weak coupling domain $\lambda <1$  the density of
the phases of  the eigenvalues
is supported by an interval $[-\phi , \phi]$ with $\phi < \pi$.

\subsec{The gauge theory in terms of nonrestricted complex matrices}

Now let us convert this model into a theory with flat measure and
regular interaction using the Laplace transform of the measure in the
fields  as in the
$U(N)$ vector model. This has been actually done in the mean field
analysis of lattice gauge theories
 \jb .
We insert the exponential representation of the $\delta$-function
\eqn\dddd{\delta (U^j_i-V^j_i)= {1 \over 2\pi i}
\int_{-i\infty}^{i \infty}dH^i_j \ e^{\bar H^i_j(U^j_i-V^j_i)}}
in the definition of the partition function for every matrix element
$(U_{\l})^j_i$ of the gauge fields. Here $\V_{\l}$ and $\H_{\l}$
are unrestricted complex matrices on which the integration is
performed with a flat measure
\eqn\mmmemem{d\V=\prod_{i,j=1}^N dV^i_jd\bar V^i_j,
\ d\H=\prod_{i,j=1}^N dH^i_jd\bar H^i_j}
For each oriented link $\l = <xy>$ only one pair
 of these fields is introduced
and the convention
\eqn\yty{
\V_{<yx>}=\V_{\l^{-1}}= \V^{\dag}_{\l}=\V^{\dag}_{<xy>},\ \ \
\H_{<yx>}=\H_{\l^{-1}}= \H^{\dag}_{\l}=\H^{\dag}_{<xy>}}
is assumed.
Denoting by
\eqn\onlin{e^{N^2F(\H)}=\int [d \U] e^{N\tr (\H^{\dag}\U+\H\U^{\dag})}}
the one-link mean-field integral
we write the partition function \discr\ in the form
 \eqn\oi {\eqalign{
\CZ=e^{N^2\CF}&= \int \prod _{\l }\Big(
d\H_{\l}d\V_{\l}
e^{-N\tr (\H^{\dag}_{\l}\V_{\l} +\H_{\l}\V^{\dag}_{\l})}\Big)\cr
&\prod_{\l} e^{N^2
F(\H_{\l})} \prod _{p}e^{N^2 S(\V_ p)}\cr}}
where $\V_ p$ denotes the ordered product of link variables along the
boundary $\p p$  of the elementary square $p$
\eqn\cads{\V_ p=\prod_{\l \in \p p} \V_{\l}.}
The measure is now flat and the integrand is regular.
The constraint \unitt\ is satisfied in the sense of mean values
\eqn\unvt{\langle (\V^{\dag}_{\l}\V_{\l}\ -\I)  ...\rangle =0}

In order to give topological meaning of the  $1/N$ expansion
we need to express the two potentials $F$ and $S_{\lambda}$
in terms of the moments $\tr(\ \cdot \  )^n, n=1,2,...,$ of their arguments.

\subsec{The potential for the $\H$-field}
The function $F$ can be determined using the Ward identity
\eqn\uniiy{ [\tr (\p /\p \H^{\dag} \p / \p \H)-N]e^{F(\H^{\dag}\H)}=0}
 which is the matrix analogue of \wwwar .
In the  large-$N$ limit this function  was calculated explicitly
by Brezin and Gross \ref\brgr{E. Br\'ezin
and D. Gross, \pl 97B (1980) 120} in terms of the eigenvalues of the
Hermitean matrix $\H^{\dag}\H$.
For sufficiently small field $\H$ it can be expanded as a series in the
momenta
\eqn\mmmi{\a_n={ \tr \over N} (\H^{\dag}\H)^n,}
\eqn\odof{ F[\a]=\sum_{n=1}^{\infty}
\sum_{k_1,...,k_n \ge 1}
f_{[k_1,...,k_n]}{\a_{k_1}
...\a_{k_n} \over n!}}
The coefficients of the series were found in the large $N$ limit
by O'Brien and Zuber \ref\objb{K.H. O'Brien and J.B. Zuber, \pl 144 B (1984)
407}
as a solution of
a system of algebraic relations
suggested originally  by Kazakov in \kaza .
These relations are equivalent to  eq. \uniiy\ written in terms of loop
variables
\eqn\uniyy{
\Big( \sum_{n\ge 1}n\a_{n-1}{\p_n \over N^2}+
\sum_{k,n \ge 1}[ (n+k+1)\a_n\a_k{\p_{k+n+1} \over N^2}+
 nk\a_{n+k-1}{\p_n \over N^2} {\p_k \over N^2} ] -1
\Big)\ e^{N^2F[\a]}=0}
where we denoted $\a_0=1, \p_n=\p / \p \a_n$.
Eq. \uniyy\
is sufficient to determine   the   $1/N$ expansion of the link-vertices
$f_{[k_1,...,k_n]}$.
In the large $N$ limit the identity \uniyy\ is equivalent to a
 system of recurrence relations \obrz
\eqn\oopol{\eqalign{
f_{[1]}&=1\cr
f_{[k_1,...,k_n]}+\sum _{j=2}^{n} k_j f_{[k_1+k_j,k_2,...,\bar k_j,
...,k_n]}
+\sum_{k=1}^{k_1 -1} \sum _{\alpha \in P(L)} f_{[k_1-k, L\backslash \alpha]}
f_{[k,\alpha]}&=0\cr}}
where $L={k_2,...,k_n}$ and $P(L)$ is the set of all subsets of $L$,
including the emply set, and the bar means omitting the argument below it.
The equations for the  weights $f_{[k]}$ of the disks \kaza
\eqn\plopp{f_{[1]}=1; \ \
f_{[n]}+\sum_{k=1}^{n-1} f_{[k]}f_{[n-k]}=0, \ n=1,2,...}
are solved by the Catalan numbers
\eqn\ihy{f_{[n]}=(-)^{(n-1)}{(2n-2)! \over n! (n-1)!}}

 \subsec{The potential for the $\V$ field}

The potential $S_{\lambda}(V)$ is defined by the analytic continuation
of the r.h.s. of \chpo\ from the hyperplane $\V=\U,\ \  \U^{\dag}\U=\I$ to
the space of all complex matrices $\V$.

Using the explicit formula for the characters \chhr\ we can express the
r.h.s. of \chpo\ as a series in  the moments
\eqn\mimmi{ \beta _n=
{ \tr \over N} \V^n , \  \beta _{-n}={ \tr \over N}
  \V^{\dag n}
;\ \ \ \  \ n=1,2,... }
The series is convergent for small $\V$ and can be represented as
exponential of another series
\eqn\asda{N^2 S_{\lambda}[\beta]=N^2 \sum_{n=1}^{\infty}
 \sum_{k_1,...,k_n \ne 0}
s_{[k_1...k_n]} {\beta _{k_1}...\beta _{k_n}
\over n!}}

In the continuum limit $\lambda \to 0$ the expansion $\asda$ takes the form
\foot{This formula resembles eq. (22) in the paper by M. Douglas \dg .
We tried to establish an exact correspondence between his
formalism and ours, but haven't succeed in this}
\eqn\ehaa{S_0(\V)=\sum _{n=1}^{\infty}{1 \over n}
 (\beta_n +\beta_{-n} - \beta_n\beta_{-n} ) = \prod _{i,j-1}^N
{1-v_i\bar v_j \over (1-v_i)^N(1-\bar v_j)^N}
}
which tends to the  $\delta$-function $\delta (\V,\I)$
 when $\V$ approaches the hypersurface $\V^{\dag}\V=\I$.

Note that the potential \asda\ for the Wilson action
has the  form $\tr (\V+\V^{\dag})$ only in the strong coupling domain.
In the weak coupling domain $\lambda <1$
 it is given by  the most
general expansion \asda\ with complicated coefficients.
 Therefore the heat kernel action is
simpler to work with in the weak coupling regime.

The lowest-order coefficients for the heat
kernel action are, in the large $N$ limit
\eqn\exx{\eqalign{
&s_{[1]}=s_{[-1]}=e^{-\lambda /2}, \ s_{[2]}=s_{[-2]}
(1-\lambda)e^{-\lambda}, \cr
 &s_{[1,1]}=s_{[-1,-1]}=
(\lambda - \lambda^2/2) e^{-\lambda},
s_{[1, -1]}=-e^{-\lambda}\cr}
}

\subsec{The $U(N)$ gauge theory \`a la Weingarten}
The interpretation of the functional integral as a sum over surfaces is
possible if the action is linear in the traces of the matrix fields.
This is achieved by the integral transformation
\eqn\eff{e^{N^2F(\H^{\dag}\H)}=
\int
\prod_{n=1}^{\infty}
(d\a_n d\ta_n
\ e^{
N^2[ { \ta_n \over n} {\tr\over N}  (\H^{\dag}\H)^n  -
{\ta_n\a_n \over n}] }) \ \ e^{ N^2 F[\a]}}
By means of another system of parameters
we represent the exponential of the plaquette  action
 in the form
\eqn\sssos{
e^{N^2S( \V)}=
\int
\prod_{n=1}^{\infty}
(d\beta _n d\tilde \beta _n d\beta _{-n} d\tilde \beta _{-n}
e^{N^2[ {\tilde \beta _n \over n} ({\tr\over N} \V^n - \beta _n )
+ {\tilde \beta _{-n}\over n}( \tr \V^{\dag n}
  -\beta _{-n})]}
) e^{N^2  S[\beta ]}
}

We therefore introduce at each link $\l$ and at each elementary square  $p$
a set of auxiliary loop variables
coupled to the moments of the matrix fields
\eqn\aaapa{\a_n (\l), \ta_n(\l); \  \
 n=1,2,...}
\eqn\vvvpv{\beta_{n}(p),\tilde \beta _n(p);\ \ n=\pm 1, \pm 2, ...}
and represent the integral \oi\ as
  a theory of Weingarten
type
 described by the partition function
\eqn\lst{\eqalign{
\CZ =e^{N^2\CF}&=
 \Bigg\langle  \int \prod _{\l }
d\H_{\l}d\V_{\l}
e^{-N\tr (\H^{\dag}_{\l}\V_{\l} +\H_{\l}\V^{\dag}_{\l})}\cr
&\prod _{\l ; n>0}e^{{N \over n} \ta_n(\l) \tr (\H_{\l}^{\dag }
\H_{\l})^n }
\prod _{p; n>0}e^{{N \over n}[ \tilde \beta _n(p) \tr \V_p^n
 +\tilde \beta _{-n}
 \tr \V^{\dag n}_p]}  \Bigg\rangle_{\ta , \tilde \beta } \cr}
}
where the Boltzmann weights of the surfaces are themselves
quantum fields and the average with respect to them is defined by
\eqn\rrrea{\eqalign{
\langle *  \rangle
_{\ta  , \tilde \beta } &=\int \prod _{\l \in \CL;n>0}
 d\a_n(\l)d\a_n(\l)
e^{-N^2 {\ta_n(\l)\a_n(\l)\over n} +F[\a(\l)]}\cr
&\prod_{p;n \ne 0} d\tilde \beta _n(p)d\beta _n(p)
e^{-N^2{\tilde \beta _n(p) \beta _n(p) \over n}+S[\beta (p)]} * \cr
  }}

The integration over the matrix fields in \lst\ will produce
an effective action which will be interpreted in the next section
as a sum over closed connected surfaces on the lattice.
The leading term in this action is proportional to $N^2$.
Therefore, in the limit $N\to\infty$ the integral over the scalar
fields \aaapa\ and \vvvpv\ is saturated by a saddle point and
these fields freeze at their vacuum expectation  values
\eqn\clls{\langle \ta_n(\l)\rangle=\ta_n , \ \ \
\langle \tilde \beta _n (p) \rangle  =
\langle \tilde \beta _{-n} (p) \rangle  = \tilde \beta _n}
This is natural to expect since the auxiliary scalar fields are coupled to
gauge invariant collective excitations of the original matrix
fields (the moments $ (\cdot)^n$)
 which, due to selfaveraging, become classical in the large $N$ limit.
The classical values of the fields $\ta_n$ can be determined
from the stationarity condition for the effective action, or,
alternatively, from the Ward identity \unvt .
Thus in the limit $N\to\infty$  we obtain a  theory of
noninteracting planar ($G$=0) random surfaces with Boltzmann
 weights determined
dynamically.
This picture is conceptually the same as in the $U(N)$ vector model.

Since we are interested only in the large $N$ limit, the fields
\aaapa\ and \vvvpv\ can be replaced with their vacuum expectation
values even before the integration over the matrix fields.
We therefore find the following expression for the free energy of
the $U(\infty)$ gauge theory
\eqn\aiaiai{\eqalign{
\CF =\lim_{N\to\infty}{1 \over N^2}\log &
\Bigg(\int
  \prod _{\l }
d\H_{\l}d\V_{\l}
e^{-N\tr (\H^{\dag}_{\l}\V_{\l} +\H_{\l}\V^{\dag}_{\l})} \cr
& \prod _{\l ; n>0}e^{{N \over n} \ta_n\tr (\H_{\l}^{\dag }
\H_{\l})^n }
\prod _{p; n>0}e^{{N \over n} \tilde \beta _n ( \tr \V_p^n
 +
 \tr \V^{\dag n}_p) } \Bigg)\cr}
}
The values $\a_n,\tilde \beta_n$ should be kept as free parameters.
The parameters  $\ta_n$ are determined by the condition of unitarity
\unvt\ and $\tilde \beta _n$ can be considered as coupling
constants defining the plaquette action.

\newsec{Branched random surfaces}
\subsec{Feynman rules in the large $N$ limit}
   Assuming that the saddle point for
the integral  \oi\ is  the trivial field  $\H=\V=0$,
we can express the free energy as a sum of vacuum
Feynman diagrams.

The Feynman rules are obtained as follows.
Inverting the bilinear part of the action we find the propagators
\eqn\connt{\langle  (H_{\l})^i_j (V^{\dag }_{\l})^k_l \rangle =
\langle  (H^{\dag}_{\l})^i_j (V_{\l})^k_l \rangle
={1 \over N}\delta^i_l \delta^{k}_j ; \
\langle   (V^{\dag}_{\l})^i_j (V_{\l})^k_l \rangle
       ={\ta_1 \over N}\delta^i_l \delta^{k}_j
}
 The vertex $\tilde \beta _{n} \tr \V^{n}_p$
 is associated with the loop
$ (\p p)^n$ going $n$ times around the boundary $\p p$
of the elementary plaquette
$p$.
We attache to  the loop $(\p p)^n$ a little planar  surface
$p_n$ called $n$-plaquette.
The $n$-plaquette $p_n$
has area $n$ (we assume that the elementary plaquette has unit area)
and a boundary going $n$ times along the boundary of the plaquette $p$.
The first three multi-plaquettes  ($n=1,2,3$) are shown in fig. 2 where for
 eye's convenience we have displaced the edges of the boundary.

The $n$-plaquette with $n>1$  has a singularity
(branch point of order $n$) containing Gaussian curvature $2\pi (n-1)$
at some point, say, at its centre.
The $n$-plaquettes will serve as building blocks for constructing the
world sheet of the lattice string.
The  Boltzmann weight of an $n$-plaquette in $\tilde \beta_n$.
The propagator $\V-\V$ has simple geometrical meaning: it glues two edges
of multiplaquettes having opposite orientations.

The absence of a $H-H$ propagator means that the $\H$ field just play the
role of glue for identifying the edges of the multiplaquettes.
Besides the simple contraction $\V-\V$ there are cyclic contractions
of $n$ pairs of oppositely oriented edges, $n=2,3,...$,
which are represented by the vertices  $\ta_n \tr (\H^{\dag}\H)^n$.
The edges are glued half-by-half in a cyclic way.
The surface obtained in this way has a singularity at the middle of
the link where a curvature $2\pi (n-1)$ is concentrated.
We will call as before this singularity a branch point of order $n$.
The Boltzmann factor associated with a cyclic
contraction of order $n$ is $\ta_n(\l)$.
The simple contraction and the first nontrivial  ones are shown in fig. 3.
\smallskip

\vskip 3 cm

$$ n=1 \ \ \ \ \ \ \ \ \ \ \ \ \ \ \ \ \ \ \ \ \ n=2
       \ \ \ \ \ \ \ \ \ \ \ \ \ \ \ \ \ \ \ \ \ n=3$$
\bigskip
\centerline{Figure 2.
Multiplaquettes of orders 1, 2, 3}

\vskip 3 cm

$$ n=1 \ \ \ \ \  \ \ \ \ \ \ \ \ \ \ \ \ \ \ \ \ n=2
       \ \ \ \ \ \ \ \ \ \ \ \ \ \ \ \ \ \  \ \ \ n=3$$
\bigskip

\centerline{Figure 3. Cyclic contractions of $n$ edges, $n$=1,2,3 }
\bigskip

The vacuum Feynman diagrams  are surfaces composed of multiplaquettes
glued together along their edges by means of cyclic contractions.
An important feature of  these surfaces is that they can
have branch points (that is, local singularities of the
curvature) not only at the sites, as it is the case in the original
Weingarten model, but also at the links and plaquettes  of the surface.
These singularities can be interpreted as processes of splitting
and joining of strings.
The weight of a closed surface is a product of the weights of its
elements times a power of $N$ which, with our normalizations, is
equal to the Euler characteristics of the surface.

The free energy of the $U(\infty)$ gauge theory is given by the sum over
all closed connected surfaces with the topology of a sphere.

The connection with the traditional Feynman-like diagrams is by duality.
Sometimes it is mire convenient to use the traditional
diagrammatical notations. Then a surface bounded by a loop
$\Gamma = [\l_1\l_2...\l_n]$ will be represented
by a planar Feynman graph with $ n$ external legs dual to the links
$\l_1,...,\l_n$.
The cyclic contraction of $2n$ is represented by a vertex with $2n$ lines
as is shown in fig. 4.

\vskip 3 cm

\centerline{Figure 4. Vertices dual to cyclic contractions}

\bigskip

Applying these Feynman rules to the Wilson loop average
$W(\Gamma)= \langle {\tr \over N} \V(\Gamma)\rangle$
we find a representation of $W(\Gamma)$ as the sum over all planar surfaces
bounded by the contour $\Gamma$.
The Boltzmann weight of a surface  is the product of
the mean values \clls\
associated with the multiplaquettes and cyclic contractions.
The local Boltzman weights can be decomposed into
factors contributing to the total area of the surface and the length of
its boundary, and factors related to the branch points.
We denote
\eqn\huy{\tilde \beta _n  = \ta_n = \ta_1^n \omega ^{(1)}_n, \ \ \
 \tilde \beta _1^n \omega _n^{(2)}}

Then the string path integral for the Wilson loop reads
\eqn\kkksk{
W(\Gamma)= e^{-m_0L[ \Gamma]} \sum_{\CS:\p \CS =\Gamma}
e^{-M_0 A[\CS]}
\Omega [\CS]}
where $L[\Gamma]$ is the length of the contour $\Gamma$,
$A[\CS]$ is the area of the surface $\CS$,
the bare string tension $M_0$ and the boundary mass $m_0 $ are
related to the original variables $\ta_1 , \tilde \beta_1$  by
\eqn\mmnm{e^{-M_0}=\ta_1^2\tilde\beta_1' \ \ \ e^{-m_o}=\sqrt{\ta_1}}
and the $\Omega$-factor is the product of the local factors
$\omega ^{(\sigma)}_n$
associated with the branch points.

The $\Omega$-factor of the surface $\CS$ containing
  $\CN^{(\sigma)}_n$
$n$-plaquettes, $n=1,2,...$,
  and  $\CN^{(1)}_n$
 cyclic contractions of order $n$ ,
$n=1,2,...$, is given by
\eqn\oomomo{
\Omega [\CS]=
\prod_{n=1}^{\infty}
 \prod_{\sigma =1,2} (\omega ^{(\sigma)}_n)^{\CN^{(\sigma)}_n}}
ant the area of the surface is
\eqn\aare{A=\sum_{n=1}^{\infty}n\CN_n^{(2)}}

\subsec{Irreducible surfaces (a miracle)}

Unlike the random walk, the two-dimensional surface can exist in
configurations with very uneven intrinsic geometry.
The typical syngularity is a ``neck'' representing an
intermediate closed string state of small connecting a ``baby universe''
with the main body of the surface.
The most singular configurations with necks can be readily removed from
the path integral of the string. Their contribution can be taken into
account by modifying the local Boltzmann weights associated with cyclic
contractions of edges.
Miraculously, the new Boltzmann weights
  become easily
calculable and do not depend neither on the dimension $D$ nor on the
choice of the one plaquette action!
This is why we will restrict the sum over surfaces to
irreducible ones which will render the
 random surface {\it Ansatz} much simpler than it its original version.

The configurations to be excluded are surfaces with necks occupying a
single link (fig. 5).
We call such a surface $reducible$ with respect to this link..

\vskip 5 cm

\centerline{Figure 5. A reducible surface}
\bigskip
In what follows  by sum over surfaces we   will understand
a sum over irreducible surfaces
 defined as follows.

\smallskip
\noindent {{\it Definition:}}
\smallskip
{\parindent =4em
\narrower
\noindent
A surface $\CS$ is {{\it reducible}} with respect to given link
$\l  \in \CL$  if it splits  into two
or more
 disconnected pieces after being cut along
this link.
A surface which is
not  reducible with respect to  any  $\l \in \CL$ is
 is called {\it irreducible}.

\par}

\subsec{Evaluation of the weights of the  cyclic contractions}

Consider  the Wilson loop average for the
 contour $\Gamma=(\l_1\l_2...)
$ with coinciding endpoints.

We will exploit the unitarity condition $\V\V^{\dag} =\I$,
applied to the loop amplitude.
It means that the Wilson average $W(\Gamma)$
will not change if a backtracking piece
$\l \l^{-1}$ is added to the contour $\Gamma$
\eqn\unilop{W(\Gamma \l\l^{-1})=W(\Gamma)}

The sum over surfaces spanning the loop $\Gamma \l\l^{-1}$ can be divided into
two pieces
\eqn\lpol{W(\Gamma\l\l^{-1})=W(\Gamma)
W(\l\l^{-1}) + W_{{\rm conn}}(\Gamma \l\l^{-1})}
The first term is the sum over all  surfaces made of two disconnected
parts spanning the loops $\Gamma$ and $\l\l^{-1}$. The  second term
contains the rest.
The constraint \unilop\ is satisfied for all loops if $W(\l\l^{-1})=1$
and $W_{{\rm conn}}(\Gamma \l\l^{-1})=0$.
But there is only one irreducible surface spanning the loop $\l\l^{-1}$;
it contains a single contraction  $\ta_1(\l)$ between  $\l$ and $\l^{-1}$.
Therefore $\ta _1 =1$.

Now consider a surface contributing to the second term $W_I$.
The
links  $\l$ and $\l^{-1}$  may be connected to the rest of the surface
by means of  the same contraction or  by  two cyclic contractions
  The condition that their total contribution is zero
is
\eqn\rscx{\ta_n + \sum _{k=1}^{n-1} \ta_k  \ta_{n-k}
=0, \ \ n=2,3,...}
This equation is illustrated in fig. 6  by means of
 the standard  graphical notations
(fig. 4).
Eq. \rscx\ is identical to the loop equation  satisfied by the
coefficients $f_n$ in the expansion \odof , which
is solved by the Catalan numbers
\eqn\iihy{ \ta_{n}=f_{[n]}=(-)^{(n-1)}{(2n-2)! \over n! (n-1)!}}
\smallskip
\vskip 3  cm

\centerline{Figure. 6. Graphical representation of the unitarity condition}

The second
 derivation
is based only on the
 the sum over surfaces for the
trivial Wilson loops
\eqn\trivv{W[(\l \l^{-1})^n] = \langle {\tr \over N} (\V_{\l}\V^{\dag}_{\l}
)^n \rangle =1, \ n=1,2,... }
For each of these Wilson loops the sum over irreducible surfaces
contains only finite number of terms, namely,
the link-vertices contracting directly the edges of the loop
$\l\l^{-1}$. For example,
\eqn\axa{W(\l\l^{-1})=\ta_1 ,W[(\l\l^{-1})^2]=2\ta_1 ^2+\ta_2, ...}
Introducing the generating functions
\eqn\genn{w(t)= \sum_{n=0}^{\infty}t^n W[(\l \l^{-1})^n]
={1 \over 1-t}, \ \ f(t)=1+\sum_{n=1}^{\infty}t^n (\ta_n)^n}
we easily find the relation \ivn
\eqn\stary{w(t)=f[tw^2(t)]}
which is solved by the function generating the Catalan numbers
\eqn\fff{f(t)={1+\sqrt{1+4t} \over 2}}

In this way, unlike the $U(N)$ vector model, the classical
values of the auxiliary fields
 {\it do not depend on the dimension of the space-time}.
The expectation value  $\ta$ in the vector model is determined by the
long wave excitations of the random walk.
Here in the gauge theory, the expectation values of the fields
$\ta_n$ are determined in purely local way, as it is clear from
their derivation.
One way to explain this difference is the local character
of  the $U(N)$ invariance in the case of the gauge theory.

On the contrary, the weights of the
multiplaquettes will depend on the dimension as
well as on  the choice of the one-plaquette action.
If wee are interested only in the large $N$ limit,
it is more convenient
 not to try to calculate them by solving the equations of motion
but just to take them as independent coupling constants.
The continuum limit (if it exists) then will be achieved along a
trajectory
\eqn\ehaaa{\tilde \beta _n= \tilde \beta _n (\lambda , D), \ \ \
\lambda \to 0}
where $\lambda$ is the coupling constant.

\subsec{String representation of the Wilson loop average.
Resum\'e}

Let us summarize what we have achieved by now.
The Wilson loop average $W(\Gamma)$
 in the $U(\infty)$ gauge theory defined on
a $D$-dimensional lattice is equal to the sum of all planar
 irreducible surfaces having as a boundary the loop $\Gamma$.
These surfaces are allowed to have branch points
  of all orders
located at the sites, links and plaquettes of the space-time lattice.
Introducing unified notations $\omega_{n}^{(k)}$
 for the weights of the  branch points associated with
the $k$-cells of the space-time lattice (sites are 0-cells, links ate
1-cells, and plaquettes are 2-cells)
where
\eqn\omom{\eqalign{
\omega^{(0)}_ n
&= 1\cr
\omega^{(1)}_n &= f_{[n]}=(-)^{(n-1)}{(2n-2)! \over n! (n-1)!}\cr
\omega^{(2)}_n &= {\tilde \beta_n \over (\tilde \beta_1)^n}
}}
we can write the sum over surfaces as
\eqn\willsl{W(\Gamma)=
\sum_{ \p \CS =\Gamma} \Omega (\CS) e^{-M_0 {\rm Area (\CS)}}}
where
the factor $\Omega (\CS)$ is a product of the weights of all branched points
of the world sheet and $M_0= - \ln \tilde \beta_1$.

\subsec{The contribution of the surfaces with folds is zero}
The weights of the branch points
 provide a mechanism of suppressing the backtracking
motion   of the strings or, which is the same, world surfaces having folds.
The  surfaces with folds have both positive and negative weights and
their total contribution to the string path integral is zero.
We have checked that in many particular cases but the general proof is missing.

It is perhaps instructive to give one example. Consider the
surface in fig. 7.

\vskip 5 cm

\centerline{Figure 7.  A piece of surface having a fold}
\bigskip

It  covers three times the interiour of the nonselfintersecting
loop $C$ and once - the rest of the lattice.
There are two specific points on the loop $C$ at which the curvature
has a conical singularity; they can be thought of as the points
where the fold is created and annihilated.
Each of these points can occur either at a site or at a link (in the
last case it is associated with a weight factor $\omega^{(1)}_2=f_{[2]}=-1$).

Let us evaluate the total contribution of all irreducible  surfaces
 distinguished by the positions of the two singular points.
Denoting by $n_0$ and $n_1 (=n_0)$
 the number of sites and links along the loop $C$,
and remembering that the a singular point located at a link has to be taken
with a weight $
f_{[2]}=-1$, we find that the contribution of these surfaces
is proportional to
\eqn\ffafa{[{n_0(n_0-1) \over 2}+f_{[2]}n_0n_1+f_{[2]}^2{n_1(n_1-1)\over 2}]
-[n_1 +2 f_{[2]}n_1] =0}
The second term on the left hand side contains the contribution of the
reducible surfaces which had to be subtracted.
A reducible surface arises when the two points are located at the extremities
of the same link ($n_1$ configurations) or when one  of the points is
a   branch point and the other is located at one of the extremities of the
same link ($2n_1$ configurations).

\subsec{Loop equations}
In ref. \ivn\ we proved that
the  loop equations
in  the Wilson lattice gauge theory
\ref\migg{
A.A. Migdal, $unpublished$ (1978);
D. F\"orster, \pl 87B (1979) 87; T. Eguchi, \pl 87B (1979) 91}
 are satisfied
by the
the sum over  surfaces \willsl .
Let us  only write here  the general formula which is derived in the same
fashion.
For any link $\l \in \Gamma$
\eqn\lopsr{\sum_{n=1}^{\infty} \sum _{p: \p p \ni \l} \tilde \beta_n
[(W(\Gamma (\p p)^n)-W(\Gamma(\p p)^{-n})]=
\sum _{\l' \in \Gamma}W(\Gamma_{\l\l'})W(\Gamma_{\l'\l})
[\delta(\l , \l')- \delta(\l^{-1} , \l ')]}
The sum on the l.h.s. goes over  the $2(D-1)$ plaquettes
adjacent with the link $\l$ and the closed loops $\Gamma_{\l \l'},
\Gamma_{\l' \l}$ in the contact term
are obtained by cutting the links $\l$ and $\l'$ and
reconnecting them in the  other possible way.

\subsec{ The trivial  $D=2$ gauge theory as a nontrivial
model of random surfaces}

The case $D=2$ was considered recently in details in \ref\az{I. Kostov,
Saclay preprint T93/050, 1993; submitted to \np B}.
For the heat kernel action the string tension $M_0$
and the  weights $\omega_n^{(2)}$ are
given by
\eqn\wowop{
M_{0}= {\lambda \over 2}; \ \
\omega_n^{(2)}= \sum _{m=1}^{n-1}\pmatrix{n\cr
m+1\cr}{n^{m-1} \over m!}(-\lambda)^m
=(1-{n(n-1)\over 2}\lambda + ...)}

Let us illustrate how the
string $Ansatz$ \willsl \ works for the  simplest
 nontrivial example of a Wilson loop (fig. 8)
 which have been calculated
previously using the Migdal-Makeenko loop equations
 \ref\kk{V.Kazakov and I. Kostov,
\np B176 (1980) 199, V. Kazakov, \np B179 (1981) 283}.
\vskip 5 cm

\centerline{Figure 8. A contour with the form of a flower
and a branched surface bounded by it }
\bigskip
The nonfolding branched surfaces spanning the loop will cover the three petals
of the flower (denoted by 1 in fig. 8 ) once, and the head (denoted by 2)
- twice. Therefore the total area will be always $A= A_1+2A_2$,
where $A_i$ is the area enclosed by the domain $i \ (i=1,2)$.
Each of these surfaces will have a branch point located at some site,
link or cell of the overlapping area 2.

Let us denote by $n_{0}, n_{1 }, n_{2 }$ the numbers
of sites, links, cells belonging to the domain 2.
Then the sum over the $\Omega$-factors due to the branch points
 reads
\eqn\omfac{\sum \Omega =
\sum_{k=0,1,2}n_{k}
\omega_2^{(k)} = (n_0-n_1+(1-\lambda)n_2)=1-n_2\lambda}
(Here we used the Euler relation $n_0-n_1+n_2 =1$.)
Therefore
\eqn\womg{W(C)=(1-A_2) e^{-{1 \over 2}(A_1+2A_2)}}

\newsec{ Strong versus weak coupling}

Let us first discuss the case of the Wilson action where
the mean field problem
is solved  exactly in the large $N$ limit \brgr .
It is well known that when $N \ge 4$ and $D=4$,
the strong coupling phase of a theory with Wilson action is separated
from the continuum limit by a nondeconfining first-order phase transition.
The assumption that $\H_{{\rm cl}}
=\V_{{\rm cl}}=0$ is a local minimum of the free energy is justified in the
strong coupling phase of the gauge theory.
In the weak coupling phase  the matrix
fields will develop vacuum expectation values.
The classical fields form an orbit of the gauge
group
\eqn\vev{\V_{<xy>}=\U_x \V_{{\rm cl}} \U^{-1}_y, \ \
\H_{<xy>}=\U_x \H_{{\rm cl}} \U^{-1}_y}
The diagram technique of sect. 2  then
 has to be modified according to the general
rules explained in \jb . The     surface elements
 will contain  free edges, and, as a consequence,
 windows will appear spontaneously on the world
sheet of the  string.
Whether these windows destroy or not the
world sheet in the continuum limit is a dynamical question.
Believing in the confinement, we expect that
 in $D\le4 $  dimensions
the windows are still not sufficiently large for that, and the world sheet
will  have in the continuum limit
 the structure of a dense network of thin strips
separating the windows (fig. 1).
 These  strips will correspond to the gluon
propagators in the standard Feynman rules.
 At the critical dimension $D=4$ the effective string
tension  of the string
 (with windows on the world sheet) should  scale with the coupling
$\lambda$  according to the
asymptotic freedom law, just as the mass in the vector model does
in $D=2$ dimensions.

In $D>4$ dimensions the world sheet of the string is
eaten by one or several  large windows and the
Wilson loop behaves as exponential of the length of the contour.

Whether this picture is true in geleral,
  can be decided by studying the
mean field problem for a general one-plaquette action.

\listrefs
%\listfigs
\bye

In $D<4$ dimensions

The weights \omom\ of the branch points consist of a
dimensionless ``topological'' term
and a term proportional to the infinitesimal area $\lambda$.
(The higher powers in $\lambda$ can be neglected in the continuum
limit.)
The term proportional to $\lambda$ is easy to interpret in the
continuum limit: it corresponds to a point-like
singularity of the intrinsic curvature
weighed by a factor $(-1)$.
The factor $\lambda$ comes from the measure over surfaces,
as we have noticed in \ivn .

The real problem consists in the interpretation of the
dimensionless weights. They seem to be associated with
puncture operators
which are pure derivatives.
These are the only singularities
 in the ``topological'' limit $\lambda =0$
of the model. In this limit  the vector potential in \i\ is restricted
to be a pure gauge,  $A_{\mu}(x)= U^{-1}(x)\p_{\mu}U(x)$.
In this ``topological'' string theory
the sum over surfaces bounded by a loop
 is always 1.

The surfaces with many necks are less rigid, i.e. have larger entropy, than
the regular surfaces and  dominate the path integral of the
lattice string and lead to a trivial critical  behaviour in $D>1$
dimensions. These configurations are ignored in the standard
world-sheet continuum formulation of the bosonic string, but they are
felt indirectly through vacuum instabilities (the appearance of
a tachyon when $D>1$).

The ``baby universes'' describe amplitudes [closed string]
$\leftrightarrow$ [vacuum] and are analogous to the tadpoles
in the standard Feynman diagram technique for point particles.
The difference is that the necks (the intermediate closed string
states)  may have various sizes  and
therefore cannot be removed by a single subtraction as in the case
of point particles.
The absence of a tachyon in the QCD string  is due to
delicate cancellations between these potentially dangerous configurations.
This is possible  at the price that  the Boltzmann weights
of the surfaces are no more positively defined.

Presumably there can exist only a finite number
of local excitations with the lowest dimension.
We have to study the classes of universality of local operators on
the world sheet to extract what is left from the puncture
operators in the continuum imit.

What we described in this talk is a game on the lattice which
may be completely irrelevant for the continuum limit.
But the very existence of such simple and beautiful geometrical
picture make us believe that the real QCD string will be found some day.

\bigbreak\bigskip\centerline{{\bf Acknowledgements}}\nobreak
The author thanks M. Douglas, D. Gross, V. Kazakov, A. Migdal,
A. Polyakov and A. Tseytlin
for stimulating discussions, and
especially  J.-B. Zuber for his critical remarks.

\listrefs
%\listfigs
\bye